# Amorphous shear bands in crystalline materials as drivers of plasticity

Xuanxin Hu[1], Nuohao Liu[1], Vrishank Jambur[1], Siamak Attarian[1], Ranran Su[2], Hongliang Zhang[3*],

Jianqi Xi[1], Hubin Luo[4, 5], John Perepezko[1], Izabela Szlufarska[1*]

1. Department of Materials Science and Engineering, University of Wisconsin, Madison, WI 53706, United States of America

2. School of Nuclear Science and Engineering, Shanghai Jiao Tong University, Shanghai 200240, PR China

3. Institute of Modern Physics, Fudan University, Shanghai 200433, PR China

4. CISRI & NIMTE Joint Innovation Center for Rare Earth Permanent Magnets, Ningbo Institute of Materials Technology and Engineering, Chinese Academy of Sciences, Ningbo 315201, PR China

5. Key Laboratory of Magnetic Materials and Devices, Ningbo Institute of Materials Technology and Engineering, Chinese Academy of Sciences, Ningbo 315201, PR China

* Corresponding authors: hlz@fudan.edu.cn (H. Zhang), szlufarska@wisc.edu (I. Szlufarska)

*Abstract*

Traditionally, the formation of amorphous shear bands (SBs) in crystalline materials has been undesirable, because SBs can nucleate voids and act as precursors to fracture. They also form as a final stage of accumulated damage. Only recently SBs were found to form in undefected crystals, where they serve as the primary driver of plasticity without nucleating voids. Here, we have discovered trends in materials properties that determine when amorphous shear bands will form and whether they will drive plasticity or lead to fracture. We have identified the materials systems that exhibit SB deformation, and by varying the composition, we were able to switch from ductile to brittle behavior. Our findings are based on a combination of experimental characterization and atomistic simulations, and they provide a potential strategy for increasing toughness of nominally brittle materials.

*Keywords*

*Main*

Shear bands are defined as thin regions with a typical thickness of several nanometers, where shear strain is localized to accommodate deformation [1, 2, 3]. Shear bands have been reported in numerous materials, including ceramics [4, 5, 6], polycrystalline metals [7, 8, 9, 10], and metallic glasses [1, 11, 12, 13]. Traditionally, shear bands in crystalline materials have been shown to emerge after severe plastic deformation, either from pre-existing defects or due to high strain-rate deformation. For example, in crystalline $B_4C$, twins [14], grain boundaries [15], and planar faults [16] were all found to be able to act as sites for the nucleation of amorphous shear bands. Amorphization within the shear band occurred as a result of accumulated damage, since high density of defects can cause the crystalline lattice to become unstable. Shear-band formation during high strain rate deformation (from ~$10^3$ to ~$10^7$ $s^{-1}$) was reported both in ceramics and in metals [9, 16, 17, 18, 19]. Formation of shear bands has been generally considered detrimental to material's toughness, because once the deformation has been localized, the shear bands can quickly form voids and initiate a crack, accelerating failure by fracture [14, 15, 20, 21].

A quite different view of shear bands has recently emerged based on studies of intermetallic $SmCo_5$ [22]. In this material, amorphous shear bands formed not as a result of defects accumulated during severe plastic deformation, but as a fundamental mechanism of accommodating plastic strain in the absence of dislocations or twins. In addition, in $SmCo_5$ shear bands did not serve as precursors to fracture, but instead continued accommodating plastic strain without cavitation, leading to a significant ductility, at least in small samples that did not have large pre-existing cracks. Similar phenomena were found in olivine deformed at 1050 °C and 1200 °C, where shear bands could form as the mechanism to accommodate shear instability rather than due to the accumulation of defects [23, 24]. The ability to control this kind of ductile shear bands has a potentially transformative impact on the design of tough structural materials. However, before the potential of amorphous shear bands can be realized, it is important to answer two overarching questions. First, when will shear bands nucleate as a fundamental mechanism of plastic strain accommodation, instead of being the final and undesirable outcome of damage accumulation? Secondly, when does this shear band enhance material's toughness, instead of nucleating voids and acting as precursors to fracture?

Here, we hypothesized that amorphous shear bands can form in materials that can easily transform between crystalline and amorphous phase. For that reason, we targeted Al-Sm systems, since they are known to have good glass forming ability (GFA) [25, 26, 27, 28]. By varying the compositions of Al-Sm systems, we found materials where shear bands lead to either ductile or brittle behavior, creating a unique opportunity to identify the criteria for void nucleation in amorphous shear bands.

We first performed molecular dynamics (MD) simulations of $Al_2Sm$ and $Al_3Sm$ samples and indeed found that the simulated sample deformed plastically through the formation of amorphous shear bands and without dislocations. We subsequently tested predictions of MD simulations experimentally by studying mechanical properties of $Al_2Sm$ and $Al_3Sm$ through nanoindentation. In our experiments, we not only found shear-band plasticity in the absence of dislocations, but also identified relations between the shear bands and the measured mechanical properties of the materials. Our experimental observations were supported by *ab initio* calculations based on the density functional theory (DFT) and additional MD simulations. Based on the combination of experimental and simulation results, we propose criteria and mechanisms to explain why different conditions are required for shear band nucleation and understand whether shear bands would lead to fracture or act as drivers to ductility. Such criteria provide effective guidance for engineering materials with high toughness.

*MD simulations of Al-Sm systems*

We have carried out molecular dynamics (MD) simulations of uniaxial compression and/or tension in two systems: $Al_2Sm$, and $Al_3Sm$ and we found that these materials deform by the formation of amorphous shear bands and without dislocations. In the following discussion we focus on Al$_2$Sm with Fd$\bar{3}$m space group and Al$_3$Sm with P6$_3$/mmc space group as these compositions were also investigated experimentally. Atomic strain from MD simulations for uniaxial tension for Al$_2$Sm and Al$_3$Sm is shown in Fig. 1a and 1b, respectively. At 13% strain, deformation is mostly localized within the thin amorphous shear bands with the highest atomic strain, highlighted by the red dashed circles. The magnified images of atomic structures in Figs. 1c and 1d reveal lattice disorder within amorphous shear bands, marked by the yellow dashed circles. The shear bands are not aligned with any crystallographic planes in either Al$_3$Sm or Al$_2$Sm.

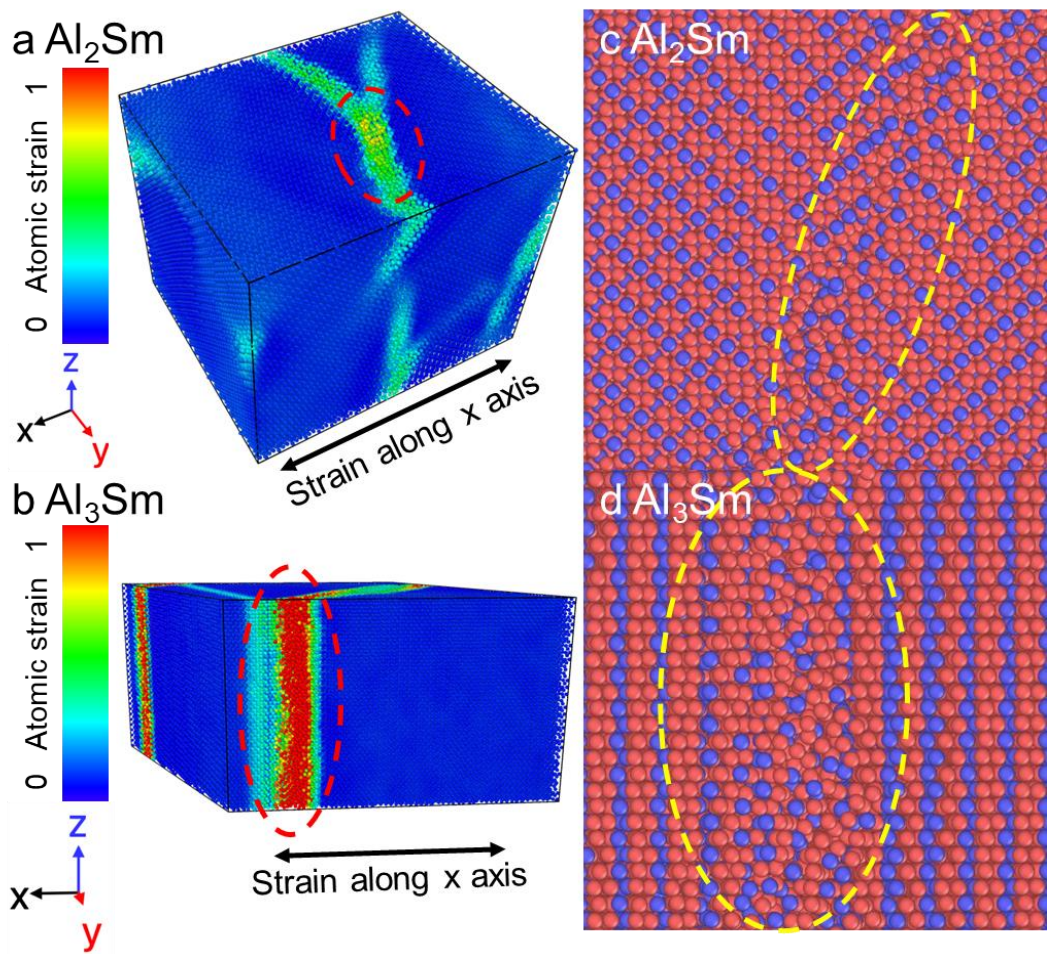

**Figure 1. Visualization of shear bands in simulated uniaxial tension.** Visualization of atomic strains in (a) Al$_2$Sm and (b) Al$_3$Sm. The regions highlighted by red dashed
4

circles are magnified to show atomic structures in the vicinity of shear bands in (c) $Al_2Sm$ and (d) $Al_3Sm$. Amorphous shear bands are highlighted by the yellow dashed circles. The snapshots are taken at 13% strain during simulations of uniaxial tension.

*Experimental observations of shear bands*

Predictions of shear bands from MD simulations were confirmed by our experiments. We first deformed $Al_2Sm$ and $Al_3Sm$ samples by means of Vicker's indentation, then performed Focus Ion Beam (FIB) lift-out followed by Transmission Electron Microscopy (TEM) analysis. The TEM images for the $Al_2Sm$ and the $Al_3Sm$ samples are shown in Fig. 2.

In both $Al_2Sm$ and $Al_3Sm$, shear bands are found to form directly under the indented region, as shown in Fig. 2a and b. The orientation of shear bands is unaligned with any specific crystallographic plane (shown in Supplementary Fig. 1). The lack of alignment of shear bands with specific crystallographic plane was also reported in $SmCo_5$ and $Sm_2Co_{17}$ [29, 30]. The High-Resolution (HR)TEM images in Fig. 2c and d demonstrate that in both $Al_2Sm$ and $Al_3Sm$, the crystalline-to-amorphous transformation was restricted to the shear-band region, whereas the surrounding region remained crystalline, as evidenced by the nano-beam diffraction patterns in Fig. 2e, f, g and h. HRTEM shows no evidence of dislocations or twins in the vicinity of the amorphous shear bands. This finding is further confirmed in the inverse Fast Fourier Transform (IFFT) images shown in Supplementary Fig. 2. We have also indexed the nano-beam diffraction patterns (see Supplementary Fig. 3), which confirmed that the crystalline regions immediately adjacent to the amorphous shear band retained their original crystal structures ($Al_2Sm$ with $Fd\bar{3}m$ space group and $Al_3Sm$ with $P6_3/mmc$ space group). That means that there is no experimental evidence that phase transformation took place prior to amorphization, which is consistent with the findings from our MD simulations. Based on the HRTEM images we have also determined the average thickness of shear bands, which is 1.148±0.04 nm and 1.17±0.06 nm for $Al_2Sm$ and $Al_3Sm$, respectively. In order to quantitatively describe shear-band formation, we estimated shear-band densities by measuring the total length of shear bands within a unit area in TEM images since the samples have almost the same thickness. The same areas of each sample are measured. The shear-band densities in $Al_2Sm$ and $Al_3Sm$ are 6.32 μm/μm$^2$ and 9.82 μm/μm$^2$, respectively.

In general, we found shear bands to be qualitatively similar in $Al_2Sm$ and $Al_3Sm$. Specifically, in both compositions shear bands could form in undefected crystalline phases during low strain-rate plastic deformation. In addition, relatively large densities of shear bands are found in both $Al_2Sm$ and $Al_3Sm$, indicating that deformation does not continue to localize within a single (or few) shear bands after their nucleation. That in turn means that failure is not going to be accelerated by continued localization of deformation, in contrast to shear band localization phenomena known in metallic glasses. However, there are quantitative differences between the two Al-Sm phases investigated here as the calculated shear-band density in $Al_3Sm$ is 55.4% higher than that in $Al_2Sm$. In order to bring insights into the underlying reasons for the above similarities and differences in shear-band behaviors, we will analyze the energetics of shear bands in $Al_2Sm$ and $Al_3Sm$ using atomistic simulations in later sections.

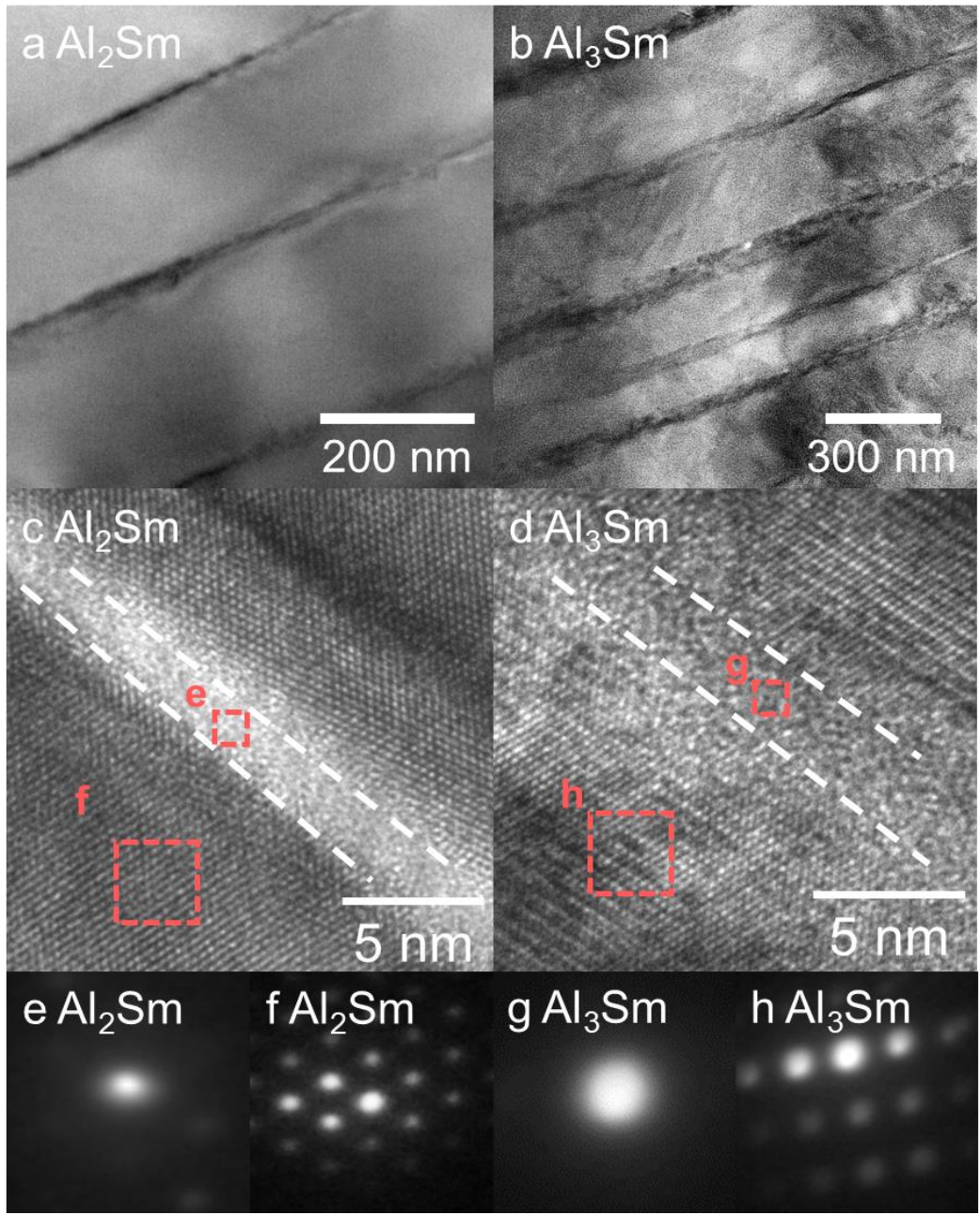

**Figure 2. Microscopy analysis of shear bands after indentation.** Bright-field Scanning TEM (BF-STEM) for (a) Al$_2$Sm and (b) Al$_3$Sm. HRTEM images for (c) Al$_2$Sm and (d) Al$_3$Sm with shear-band regions marked by the white dashed lines. Nano-beam diffraction patterns for related marked regions in the HRTEM images: (e) amorphous shear-band region in Al$_2$Sm, (f) crystalline surroundings in Al$_2$Sm, (g) amorphous shear-band region in Al$_3$Sm and (h) crystalline surroundings in Al$_3$Sm.

Because indentation can suppress cracking even in brittle materials, the ability of the materials to accommodate plastic deformation by shear-band formation was additionally evaluated in micro-pillar compression tests. We found that in this test, $Al_2Sm$ and $Al_3Sm$ exhibited quite different mechanical behaviors. Specifically, all of the 15 micropillars of $Al_2Sm$ that were tested fractured after compression. A representative example is shown in Fig. 3a-c. It is worth noting that, prior to fracture, shear bands did form in this sample, as indicated by serration in the strain-stress curve and the morphology highlighted by the red dashed circles in Fig. 3b. The fractured micro-pillar in Fig. 3c suggests that $Al_2Sm$ exhibits limited ductility even with shear-band formation. In contrast, analogous tests on $Al_3Sm$ showed substantial ductility. Specifically, we tested 15 micro-pillars of $Al_3Sm$ and found that a micro-pillar with a nominal length of 6 μm could experience at least 25% engineering strain without fracture during compression, providing strong evidence for a good ductility in the absence of large pre-existing flows (see Fig. 3d). The morphology of the deformed micro-pillar (Fig. 3e) indicates that there are many shear bands formed during compression, which is consistent with the serration observed in the strain-stress curve (Fig. 3d). Furthermore, micro-pillars of $Al_3Sm$ with different lengths could accommodate as large as 40% engineering strain without fracture (see Supplementary Fig. 4).

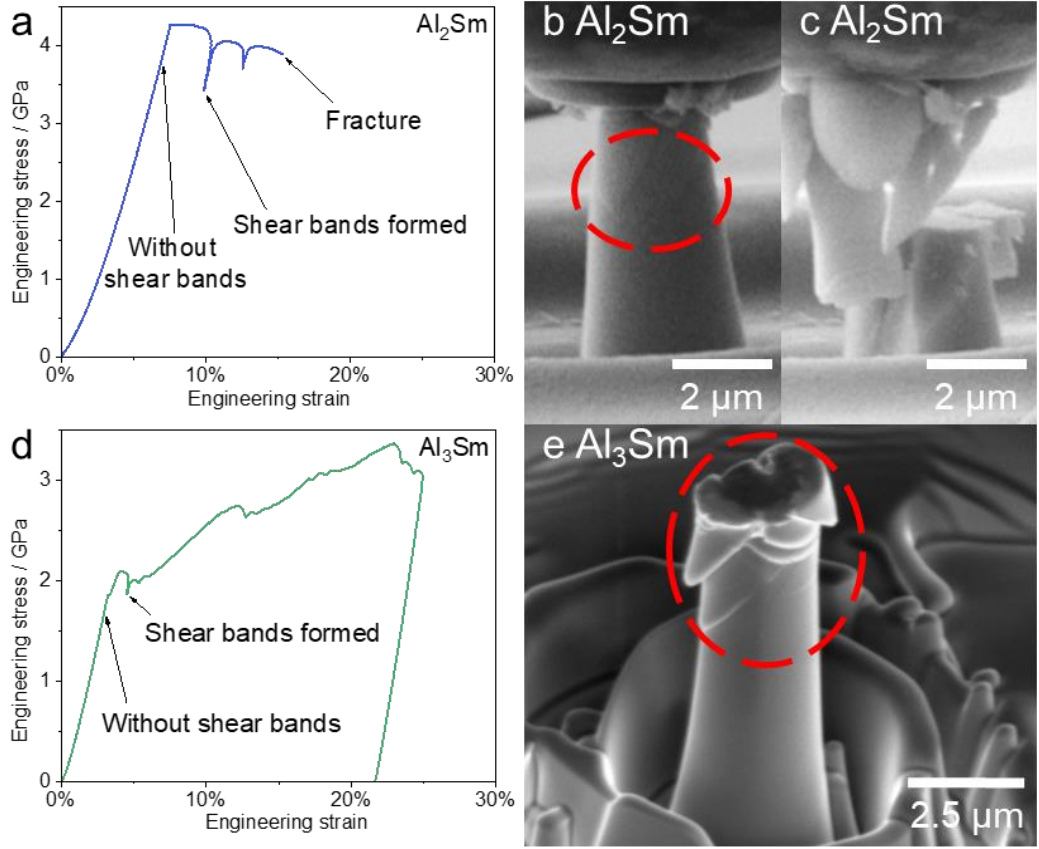

**Figure 3. Micro-pillar compression tests of Al$_2$Sm and Al$_3$Sm.** Micro-pillar compression tests: (a) (Engineering) strain-(Engineering) stress curve for Al$_2$Sm; SEM image for Al$_2$Sm (b) with shear bands and (c) after fracture; (d) (Engineering) strain-(Engineering) stress curve for Al$_3$Sm; (e) SEM image for Al$_3$Sm with shear bands after compression. The morphologies related to shear-band formation are highlighted by the red circles.

*Shear-band properties determined from atomistic simulations*

We hypothesize that the difference in energies between crystalline and amorphous phases plays an important role in how easily amorphous shear bands could nucleate in the absence of pre-existing defects. To determine the relationship between energy changes from crystalline to amorphous phases and shear-band formation, energies of different phases of Al$_2$Sm and Al$_3$Sm are calculated by DFT. As shown in Fig. 4a, the energy changes during crystalline-to-amorphous transformation for Al$_2$Sm and Al$_3$Sm are 0.2445 eV/atom and 0.1963 eV/atom, respectively. These energy changes are much smaller than that in Si, B$_4$C, and SiC, which require at least 0.4 eV/atom change in energy during crystalline-to-amorphous transformation [31, 32, 33]. In Fig. 4a, we have also included the data for SmCo$_5$, where shear bands were found to form readily inside the undefected crystalline grains [20, 22]. Fig. 4a shows that the transformation energies in different materials can be separated into two groups with the transition threshold between 0.35~0.4 eV/atom. Materials that form amorphous shear bands as a primary deformation mechanism have transformation energies lower than this threshold and materials where shear bands form as a result of accumulated damage have energies higher than the threshold. In addition, it is interesting to compare the transformation energies for Al$_2$Sm and Al$_3$Sm. The lower energy cost of amorphization in Al$_3$Sm is consistent with the higher density of shear bands observed experimentally in this sample relative to Al$_2$Sm, further suggesting that the amorphization energy is indicative of the propensity for the formation of amorphous shear bands in crystalline samples.

Traditionally, the primary mechanism to accommodate plastic deformation is dislocation motion. However, dislocations were not observed in our Al-Sm samples, either in MD simulations or in experiments, which suggests that the energy to nucleate and move dislocations in Al-Sm systems is relatively high. Toward this end, we used DFT calculations to determine the generalized stacking fault energies (GSFE) on different potential energy surfaces of rigid slip in Al$_2$Sm and Al$_3$Sm, which are correlated with the propensity for dislocation nucleation and glide [34, 35]. The energy maxima on GSFE surfaces are 1255 mJ/m$^2$ (0.0783 eV/Å$^2$) and 1116 mJ/m$^2$ (0.0697 eV/Å$^2$) for Al$_2$Sm and Al$_3$Sm, respectively. These values are one to two orders of magnitude higher than GSFE maxima in metal materials with similar crystal structures that exhibit dislocation plasticity. Although in Al$_3$Sm partial slip is possible on prismatic planes, a complete slip is energetically unfavorable and if the partial slip occurs, it was found in simulations to transform into an amorphous shear band. Details are shown in Supplementary Fig. 5, Supplementary Fig. 6, and Supplementary Table 1.

As shown in Fig. 3 and discussed earlier, during micro-pillar compression Al$_2$Sm failed in a brittle manner, whereas Al$_3$Sm showed significant ductility. We hypothesize that in the presence of shear bands, propensity for fracture might be associated with the difference in the density between amorphous and crystalline phases. Larger density differences could result in volume mismatch and large local strain in the shear bands, leading to subsequent nucleation of voids and of cracks.

Toward this end, we calculated the densities of different phases of $Al_2Sm$ and $Al_3Sm$ by DFT, listed in Fig. 4b. The density changes from crystalline to amorphous phases in $Al_3Sm$ and $Al_2Sm$ are -3.00% and -6.73%, respectively. The smaller density change in $Al_3Sm$ is consistent with a smaller change in the coordination number during amorphization, with the value of -7.06 % for $Al_3Sm$ and -17.06 % for $Al_2Sm$ and (the corresponding radial distribution functions are shown in Supplementary Fig. 7). The density difference in $Al_2Sm$ is comparable to the very large values reported in ceramics [14, 15, 30, 36, 37, 38], which are also shown in Fig. 4b. The large density changes in the shear-band regions in Si, SiC and $B_4C$ have been previously invoked as one of the hypotheses for why voids nucleate in the shear band regions [14, 15, 39, 40, 41]. However, in the absence of examples where shear bands do not nucleate voids, this hypothesis was difficult to test. Here, we are able to demonstrate this trend by considering materials that do and that do not nucleate voids, including two different compositions in a single materials system (Al-Sm). Specifically, the more brittle $Al_2Sm$ experiences a larger density change, whereas $Al_3Sm$ with a smaller density change exhibits ductility. These results suggest that materials where shear bands enhance ductility (or toughness for macroscale samples) instead of being precursors to crack formation are those with small volume/density change during amorphization.

The difference in the propensity to nucleate voids in $Al_2Sm$ and $Al_3Sm$ was confirmed in our MD simulations of uniaxial tension. Figs. 4c and 4d show atomic strain of $Al_2Sm$ and $Al_3Sm$, respectively. The curves in Fig. 4e represent the total volume of voids nucleated during the uniaxial tension, showing that a large volume of voids nucleate in $Al_2Sm$ after 13 % of true strain whereas $Al_3Sm$ does not form any voids. We have also calculated the fraction of Kasper clusters, which are close-packed atomic structures shown to correlate with local stiffness and yield resistance [42]. The fractions of Kasper clusters in the amorphous shear-band regions of $Al_2Sm$ and $Al_3Sm$ are 51.3 % and 28.5 %, respectively, which is consistent with the higher ductility of shear bands in $Al_3Sm$ (Voronoi indices and Kasper clusters are shown in Supplementary Fig. 8).

During deformation, the formation of shear bands competes not only with dislocation nucleation but also with brittle fracture. To determine the propensity for brittle fracture in Al-Sm system, we calculated cleavage energies in these systems, especially along the most close-packed planes, and compared these energies to the formation energy of amorphous shear bands. Lower shear-band formation energy indicates that shear-band formation is more energetically favorable than fracture. Since cleavage energies are determined per unit area of a surface, we express the formation energy of shear bands in the same units. Specifically, we calculate the potential energies of atoms within the shear-band region before and after amorphization, and then the energy difference is divided by the area of the shear band. The thicknesses of the shear band are assumed to be 1.5 nm and 1.2 nm for $Al_2Sm$ and $Al_3Sm$, respectively, based on the values measured in the HRTEM images above. The results are shown in Fig. 4f. Interestingly, in $Al_2Sm$ the shear-band formation energy (0.159 eV/ $Å^2$) is comparable to the cleavage energy for (111) plane (0.156 eV/ $Å^2$). It means that cleavage fracture and shear-band formation are almost equally likely to form in $Al_2Sm$ during deformation, consistent with the experimental results that the micro-pillars of $Al_2Sm$ fractured with shear bands formed in compression tests. In contrast, in $Al_3Sm$, the cleavage energies for plane (0001) and $(10\bar{1}0)$ are 0.165 eV/ $Å^2$ and 0.151 eV/ $Å^2$, respectively, about 50% larger than the shear-band formation energy of 0.110 eV/ $Å^2$. Higher cleavage energies in $Al_3Sm$ suggest that cleavage fracture is not energetically preferable. The high cleavage energy in $Al_3Sm$ combined with the small density change during crystalline-to-amorphous transformation are consistent with our observation that amorphous shear bands in $Al_3Sm$ can accommodate large plastic deformation without nucleating a crack during micro-pillar compression tests.

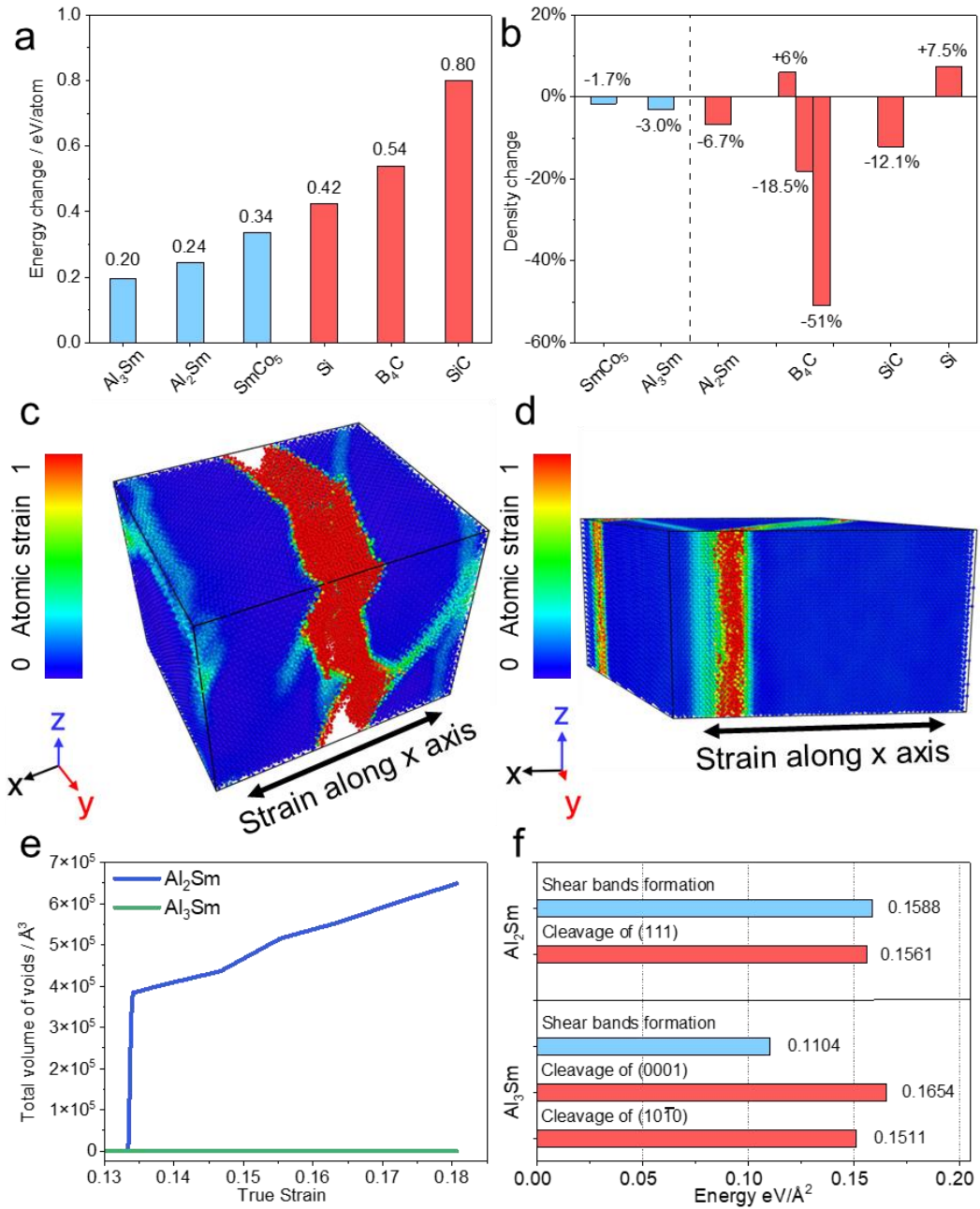

**Figure 4. Key properties to determine the shear-band behaviors.** (a) Energy difference between crystalline and amorphous phases for $Al_2Sm$, $Al_3Sm$, $SmCo_5$ [22], Si [43], $B_4C$, and SiC [31, 32, 33]. The blue and red bars, represent shear-band formation in the absence of pre-existing damage and as a result of such damage, respectively. (b) Density changes from crystalline to amorphous phases of $Al_2Sm$, $Al_3Sm$, $SmCo_5$, and a few other materials where shear bands are known to nucleate cracks [14, 15, 30, 36, 37, 38]. The blue and red bars represent materials where shear bands lead to ductility and fracture, respectively. Visualization of atomic strains in (c) $Al_2Sm$ (with voids) and (d)

Al$_3$Sm (without voids) at 20 % uniaxial tensile strain. (e) Total volume of voids nucleated in Al$_2$Sm and Al$_3$Sm during uniaxial tension. In Al$_2$Sm voids began to nucleate at around 13.5% uniaxial tensile strain. (f) Comparison of shear-band formation energies and cleavage energies for Al$_2$Sm and Al$_3$Sm, with assumed thickness of 1.5 nm and 1.2 nm, respectively.

*Mechanical properties of Al$_2$Sm and Al$_3$Sm*

If shear bands are the primary mechanisms of deformation, it is important to determine how shear bands contribute to the strength of materials. Micro-pillar compression experiments generate engineering stress-strain curves and the trend in engineering stress cannot be used directly to determine if the material work hardens or work softens. Therefore, we measured the elastic Young's modulus and hardness of Al$_2$Sm and Al$_3$Sm as a function of nanoindentation depth by applying the continuous stiffness test with fixed peak loading of 18,000 μN. As shown in Fig. 5a, the Young's modulus does not have a strong depth dependence and it is approximately 160 GPa and 115 GPa for Al$_2$Sm and Al$_3$Sm, respectively. In contrast, hardness depends on the indentation depth and shows that both Al$_2$Sm and Al$_3$Sm undergo softening in the plastic regime (Fig. 5b). For Al$_2$Sm, hardness changes from ~10.5 GPa (at 105 nm) to 9.25 GPa (at 325 nm). For Al$_3$Sm, hardness changes from 8.75 GPa to 7.8 GPa for the same range of indentation depths. This plastic softening is consistent with the STEM images in Fig. 2a and b, which show mostly parallel shear bands with no evidence that the shear bands are interacting or arrest each other's propagation.

It is interesting to compare the modulus and hardness of Al$_2$Sm and Al$_3$Sm. The higher stiffness of Al$_2$Sm is consistent with the lower (more negative) cohesive energy of this phase [44] as compared to Al$_3$Sm with respect to the same chemical potentials. The larger hardness of Al$_2$Sm is indicative of a higher resistance to plastic flow during deformation and is consistent with our experimental observation of a lower density of amorphous shear bands in Al$_2$Sm than in Al$_3$Sm. A lower density of shear bands indicates a higher resistance to shear band nucleation and therefore a higher resistance to plastic flow.

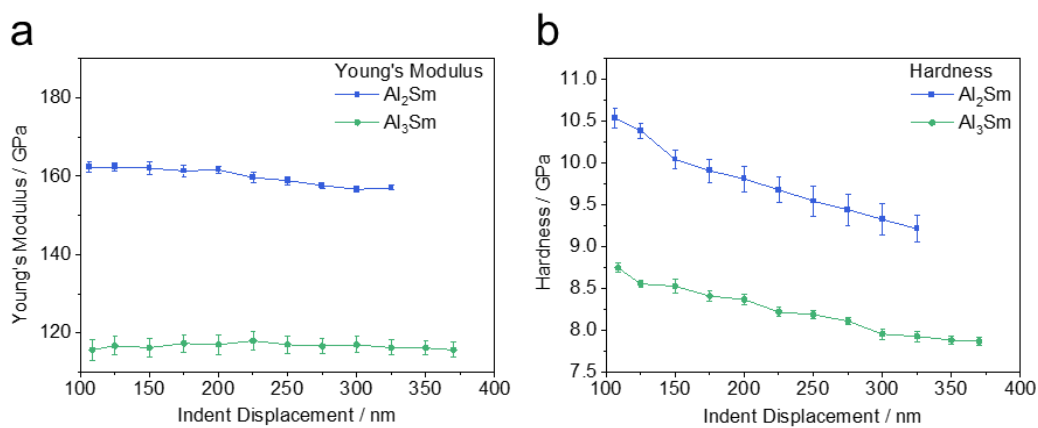

**Figure 5. Measured mechanical properties of Al$_2$Sm and Al$_3$Sm.** (a) Young's modulus and (b) hardness measured as a function of indentation depth for Al$_2$Sm and Al$_3$Sm. The error bar is the standard error of the mean (s.e.m.) with n = 10 for Al$_2$Sm and n = 8 for Al$_3$Sm. Detailed error bar values are shown in Supplementary Table 2.

*Conclusions*

Based on a combination of experimental characterization and *ab initio* DFT calculations, we have addressed two main aspects on shear-band deformation: when do shear bands nucleate as the primary mechanism of strain accommodation (as opposed to amorphization due to accumulated damage) and when do shear bands lead to void nucleation and fracture? In terms of nucleation, we demonstrated that the key criterion for shear bands to nucleate without pre-existing defects is the similarity in energies between crystalline and amorphous phases. Secondly, we provided evidence that small density changes (thus small volume change) during the transformation from crystalline to amorphous phases suppress void nucleation and allows the shear bands to be a driver of plasticity and ductility, instead of a precursor to fracture. Finally, we found that in materials that deform by shear-band plasticity, the formation of amorphous shear bands is more energetically favorable than cleavage.

Understanding when shear bands form and when they drive plastic deformation provides guidance for the design of materials with increased toughness. While brittle materials may ultimately fail in a brittle manner because of large pre-existing cracks, engineering "ductile" shear bands into such materials can increase toughness because it provides a potential mechanism for the dissipation of mechanical energy. More studies are also needed to confirm generality of the criteria identified in this paper to all crystalline materials. It will be also interesting to explore whether other (thermodynamic) criteria proposed to identify materials with good glass forming ability [45, 46] can be applied to predict the propensity of materials to form amorphous shear bands.

**Acknowledgment**

The authors gratefully acknowledge financial support from the Army Research Office Grant # W911NF2110130.

**Contributions**

I.S. directed the project. X.H. and I.S. conceived the idea. X.H. performed the transmission electron microscopy experiments and data analysis. X.H. performed the micro-pillar compression tests and date analysis. V. J., N.L. and H.L. performed the molecular dynamics simulations and data interpretation. S.A. and J.X. performed the density functional theory simulations and data interpretation. X.H. performed the continuous stiffness tests and data analysis. X.H. and R.S. performed the sample synthesis and annealing process with J.P.'s help. X.H. and H.Z. prepared the focused ion beam transmission electron microscopy samples. X.H. and H.Z. prepared the micro-pillars through focused ion beam system. X.H. and I.S. interpreted all the data, with input from all authors. X.H. and I.S. co-wrote the manuscript with input from all authors.

**Competing Interests Statement**

The authors declare no competing interests.

*Methods*

*Classical MD simulations*

Classical molecular dynamics (MD) simulations were carried out with LAMMPS [47] software package using Embedded Atomic Model (EAM) potential and Finnis and Sinclair (FS) potential. The EAM potential used in $Al_2Sm$ has previously been validated and used to model Al-Sm metallic glasses [28], and the FS-type semi-empirical potential used in $Al_3Sm$ simulation has been validated in Al-rich alloys and Al-Sm metallic glasses [48, 49]. For $Al_2Sm$, the simulations were performed with a single crystal consisting of 222,264 atoms. For $Al_3Sm$, the single-crystal structure for simulation consists of 21,600 atoms. For both cases, the atoms were first assigned random velocities from a Gaussian distribution and equilibrated in the constant pressure system (NPT) at 300 K for 250 ps. The sample was then subjected to uniaxial tension at a strain rate of $10^8$ $s^{-1}$ at 300 K under the NPT ensemble. OVITO [50] software was used to analyze the evolution of the structure of the $Al_2Sm$ crystal during deformation and for the calculation and visualization of atomic strains.

For the void detection in $Al_2Sm$, the method involved the construction of the uniform dummy particles on the lattice sites throughout the sample and deleting the particles overlap with existing atoms in the sample [51]. After this process, any remaining dummy particles will effectively fill the unoccupied space in the structure. Voronoi volume of each dummy particles is calculated and summed together, to get the overall volume of the voids existing in the structure. The construction of the dummy particle lattice and the deletion of the overlapping particles were performed by LAMMPS. The Voronoi volume is calculated using the VORONOI package in LAMMPS, which uses the open source voro++ package.

*DFT calculations in Al$_2$Sm and Al$_3$Sm*

All DFT calculations are performed with the Vienna Ab-Initio Simulation Package (VASP) [52] using the projector augmented wave (PAW) method [53]. The exchange-correlation is treated in the generalized gradient approximation (GGA) as parameterized by Perdew, Burke, and Ernzerhof (PBE) [54]. Spin polarization was considered in our simulations.

To prepare a model of glassy alloys for the DFT calculations, we first create glassy Al$_2$Sm and Al$_3$Sm systems with 192 and 216 atoms, respectively, using classical MD simulations of melting and quenching, with the same quench rate of 1 K/ps; the setting of the classical MD simulations is discussed elsewhere [55]. The resulting configurations are further equilibrated in *ab initio* molecular dynamics at 1300 K for 1 ps in the constant volume – constant temperature (NVT) ensemble with the Nose-Hoover thermostat and a time step of 3 fs [28, 55]. Afterward, the systems are optimized in the DFT calculations. All DFT calculations are performed with a cutoff energy of 500 eV for the plane-wave basis set and with spin-polarized conditions. The k-point mesh is set to 2×2×2. All lattice parameters and atomic coordinates are fully relaxed until the forces on individual atoms are smaller than $1\times10^{-2}$ eV/Å.

The generalized stacking fault energy (GSFE) curves and cleavage energies of the surfaces studied here were calculated by modeling supercells of Al$_2$Sm and Al$_3$Sm with a 15 Å vacuum above the surfaces. The number of atomic layers in each supercell is different (between 16 to 48 layers) and this number is determined based on the size of the unit cell and convergence tests of the surface energies with errors less than $10^{-2}$ J/m$^2$.

To calculate the GSFE curve for each slip system, we moved the atoms in the top half of the supercell along a specific direction in 10 equal steps until the atoms reached the positions of their periodic images. The only exception is for $(0001)[01\bar{1}0]$ system where due to the longer path we used 20 steps. At each step, the supercell size was fixed, and the atoms were only allowed to relax their positions along the direction normal to the plane for which the GSFE was calculated. GSFE curves were plotted by calculating the differences between the energies at each step and the energy of the first step. The cleavage energy was calculated using the following equation:

$$E_{cv} = \frac{E_{slab} - n \times E_{bulk}}{A}$$

where $E_{cv}$ is the cleavage energy, $E_{slab}$ is the energy of the slab (supercell), $E_{bulk}$ is the energy of each unit cell in its bulk form, $n$ is the number of unit cells used to construct the supercell, and $A$ is the cross-sectional area of the supercell (half of the total exposed area). A k-point mesh of 4×4×1 was used. PAW-PBE potentials, which were used in this study are Al ($3s^23p^1$) and Sm_3 ($4d^{10}6s^1$). The f electrons of Sm are not considered in the valence states as they are localized and do not have a significant effect on bonding that determines the energies.

*Experimental synthesis*

Bulk Al$_2$Sm and Al$_3$Sm ingots were synthesized from Al (purity 99.999%) and Sm (purity 99.9%) pieces by arc-melting and then cut into slices with a thickness of about 4 mm. Subsequent annealing was applied for crystallization. The sliced samples were annealed under vacuum at 600 °C for up to 50 h (labeled as Al$_2$Sm-600°C-annealing time and Al$_3$Sm-600°C-annealing time). The crystallinity and phase purity of annealed samples was examined by X-Ray Diffraction (XRD) (shown in Supplementary Fig. 9). No impure phase was identified in any of the Al$_2$Sm and Al$_3$Sm samples. The average grain size and internal stress were characterized by Electron Backscatter Diffraction (EBSD). Al$_2$Sm samples were found to have average grain sizes of 400 μm, 600 μm, and >1000 μm after annealing at 600 °C for 6 h, 12 h, and 50 h, respectively. Al$_3$Sm samples had average grain sizes of 50 μm, 100 μm, and 200 μm after annealing at 600 °C for 6 h, 12 h, and 50 h, respectively. Al$_2$Sm-600°C-50h and Al$_3$Sm-600°C-50h were selected for further investigation because of their highest level of crystallinity and the largest average grain size. Henceforth, we will refer to the Al$_2$Sm-600°C-50h and Al$_3$Sm-600°C-50h samples as Al$_2$Sm and Al$_3$Sm, respectively.

*Characterization methods*

X-Ray Diffraction (XRD) was performed by Bruker D8 Discovery high resolution X-ray diffractometer with 2D detector using Cu-K$_α$ radiation with a wavelength of 1.54184 Å at 50 kV voltage and 1000 μA current. To eliminate the influence of crystal preferred orientation, the sample stage rotated at a speed of 1 rpm while testing.

Electron Backscatter Diffraction (EBSD) for identifying the average grain size and internal stress was performed using the electron beam of Focused Ion Beam (FIB) FEI Helios PFIB G4 (FEI, Inc.) with a voltage of 30 kV and a current of 51 nA. The Vicker's indentation was carried out with 100 gf loading force and 10 s holding time, using a diamond square pyramid tip with an angle between the opposite surfaces of 136°.

The standard lift-out techniques based on the FIB system were applied to prepare samples for Transmission Electron Microscopy (TEM) analysis. A 3 μm Pt protective layer was deposited on the wanted indentation surface by two steps to protect the surface from ion-beam damage during lift-out process: (i) a Pt layer with 1 μm thickness was directly deposited on the indentation surface by a 5 kV and 0.8 nA low-energy electron beam, to avoid damage from subsequent high-energy ion beam deposition; (ii) another 2 μm Pt layer was deposited on the same region by a 12 kV high-energy ion beam. In order to obtain the final TEM sample with a thickness of around 100 nm, the lift-out samples were thinned by a two-step process: (i) a speed-up process using 30 kV high-energy ion beam with currents of 4 nA, 1 nA, 0.3 nA and 0.1 nA in sequence; (ii) a cleaning process using 5 kV low-energy ion beam with a current of 0.1 nA. Two lift-out samples were prepared for each of the two compositions, Al$_2$Sm and Al$_3$Sm. The lift-out regions were almost the same, as shown in the sketch diagram (Supplementary

Fig. 10). FEI Tecnai F-30 with field emission gun was used for TEM and high-resolution TEM (HRTEM), to analyze the shear bands on a microscopic scale. Analysis of TEM images, including Fast Fourier Transform (FFT) and Inverse Fast Fourier Transform (IFFT), are performed by DigitalMicrograph Software. Brightness and contrast are adjusted to highlight features (e.g., amorphous shear bands) in the images.

To provide a detailed analysis of the amorphous shear bands and local microstructure, we conducted nano-beam diffraction on two focused ion beam (FIB) lift-outs extracted from Vicker's-indented $Al_2Sm$ and $Al_3Sm$ samples. Nano-beam diffraction involves recording an electron diffraction pattern while scanning a small, parallel electron probe over the sample. Automated diffraction indexing algorithms are used to match the recorded diffraction patterns against theoretically derived templates, enabling simultaneous determination of both phase and orientation. The nano-beam diffraction was performed using a Tecnai TF30 transmission electron microscope equipped with a Nanomegas ASTAR scanning precession electron diffraction (SPED) system, operated at 300 keV. To optimize resolution and minimize the beam size, we set the spot size to 10 and the condenser aperture to 1. The diffraction patterns were acquired at 25 frames per second with a step size of 0.2 nm and a precession angle of 0.4 degrees.

The micropillar samples were prepared by FIB and they had nominal lengths ranging from 2.5 μm to 6 μm and diameters of around 1.8 μm on the top. The micropillar compression experiments were performed in a Zeiss Leo 1550VP Scanning Electron Microscopy (SEM) equipped with a Hysitron PI85 SEM Pico Indenter. A diamond-made flat-bottomed conical nanomechanical probe with a nominal flat diameter of 5.1 μm and a nominal cone angle of 60 ° was applied to do the compressions. The maximum displacement of the indenter varied from 500 nm to 2000 nm, based on different lengths of micropillars. The indenter kept a constant loading and unloading rate of 50 nm/s.

The continuous stiffness tests were performed using Hysitron TI 950 TriboIndeter equipped with a standard diamond Berkovich tip. The preset program of the instrument for continuous stiffness tests was used at a constant strain rate and a constant frequency of 220 Hz. The peak loading in our experiment was set as 18000 μN. The tests were carried out 10 times for each sample as parallel experiments.

*Data availability*

Experiments data that support the results of this work are available in the following link:

https://drive.google.com/drive/folders/16zsS_jBFdIwUaJo2euQgYRjgxYpSYfwT?usp=sharing

Simulation data that support the results of this work are available from the corresponding authors upon reasonable request.

*Methods-only references*

# Supplementary Information

# Amorphous shear bands in crystalline materials as drivers of plasticity

Xuanxin Hu[1], Nuohao Liu[1], Vrishank Jambur[1], Siamak Attarian[1], Ranran Su[2], Hongliang Zhang[3*],
Jianqi Xi[1], Hubin Luo[4, 5], John Perepezko[1], Izabela Szlufarska[1*]

1. Department of Materials Science and Engineering, University of Wisconsin, Madison, WI 53706, United States of America

2. School of Nuclear Science and Engineering, Shanghai Jiao Tong University, Shanghai 200240, PR China

3. Institute of Modern Physics, Fudan University, Shanghai 200433, PR China

4. CISRI & NIMTE Joint Innovation Center for Rare Earth Permanent Magnets, Ningbo Institute of Materials Technology and Engineering, Chinese Academy of Sciences, Ningbo 315201, PR China

5. Key Laboratory of Magnetic Materials and Devices, Ningbo Institute of Materials Technology and Engineering, Chinese Academy of Sciences, Ningbo 315201, PR China

* Corresponding authors: hlz@fudan.edu.cn (H. Zhang), szlufarska@wisc.edu (I. Szlufarska)

## Contents

## 1. Supplementary simulation method

The coordination number getting from the simulation is to count the number of nearby atoms around the central atom within the cutoff distance, and get the distribution of the fraction of atoms with different coordination numbers. According to the radial distribution function, a 3.6 Å cutoff distance is used for $Al_3Sm$ crystalline phase, where all the bonded atoms to the central atoms are taken into account. We keep the same cutoff distance in the amorphous phase to compare the density change. Similarly, a 4 Å cutoff distance is chosen for both crystalline and amorphous phases in $Al_2Sm$. The crystalline phases are the status after the 250ps NPT equilibrium before tension is applied, and the amorphous phases are the amorphous shear band area.

Voronoi analysis is a common method to analyze the atomic structure in the amorphous phase, by dividing the space into Voronoi polyhedral around every particle. The Voronoi indices are the characteristic signature of a particle's coordination structure and the topology of the Voronoi polyhedron. For the index $(n_1, n_2, n_3, n_4, …)$, $n_i$ is the number of polyhedron faces with *i* edges/vertices, and the first two indices are usually omitted since faces with less than three edges do not exist. The Voronoi indices for $Al_2Sm$ and $Al_3Sm$ are outputted by OVITO, followed by further processing to get the fraction of every index. The Kasper polyhedra where the coordination number ranges from 6 to 17 are considered to have closer packs than non-Kasper polyhedra. The proportion of indices belonging to Kasper polyhedral in $Al_2Sm$ and $Al_3Sm$ are calculated respectively.

The partial slip simulation was performed in the single-crystal structure $Al_3Sm$ consisting of 21,600 atoms, equilibrated in NPT ensemble at 300 K for 250 ps and then subjected to uniaxial tension along X direction at a strain rate of $10^8$ $s^{-1}$ at 300 K under the NPT ensemble. Some partial slips are found before the shear band formation. The slip distance is calculated based on the distance change with nearby atoms before and after slip happened.

## 2. Figures and Tables

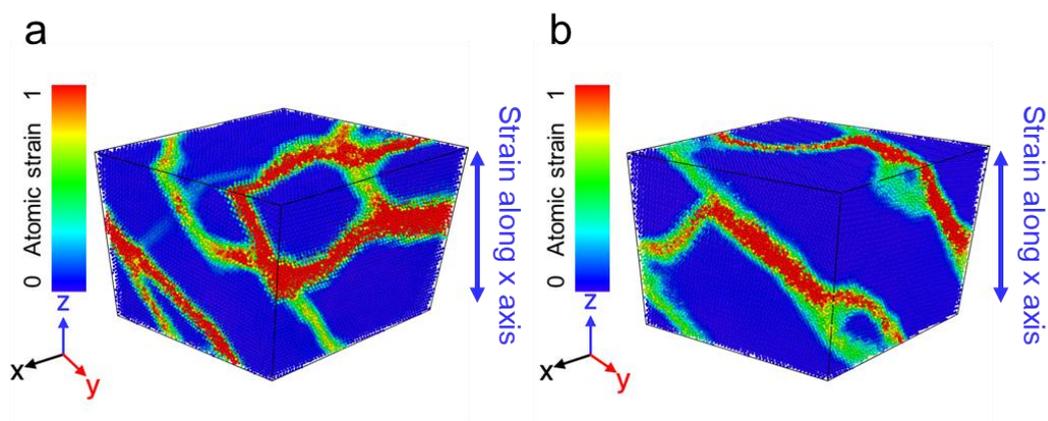

**Supplementary Figure 1.** (a) and (b) Visualization of atomic strains in $Al_3Sm$ with different random seeds in Lammps to generate the Gaussian distribution of velocities for two simulations cases, while keeping all the conditions (including the temperature and boundary conditions) the same.

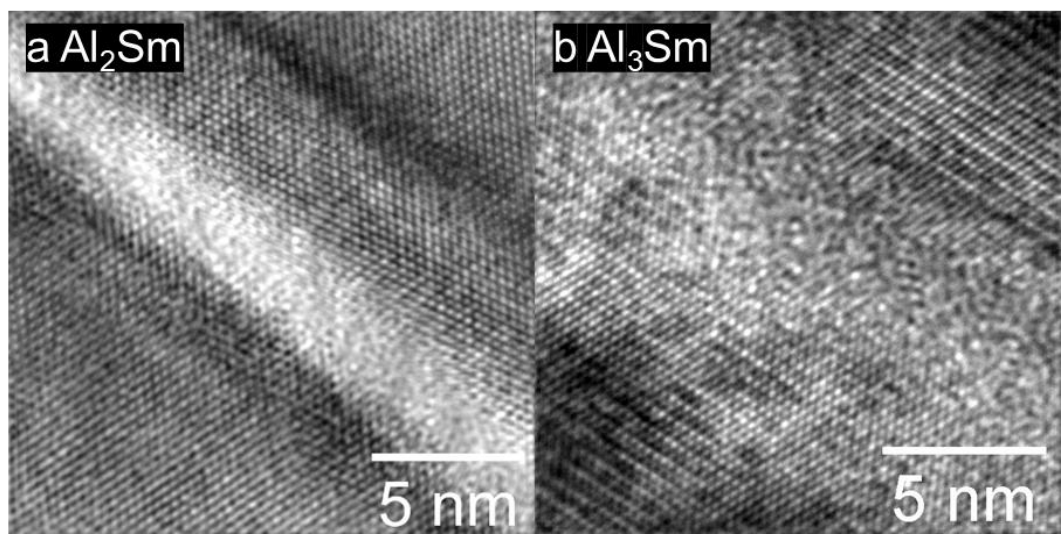

**Supplementary Figure 2.** The inverse Fast Fourier Transform (IFFT) images for (a) $Al_2Sm$ and (b) $Al_3Sm$.

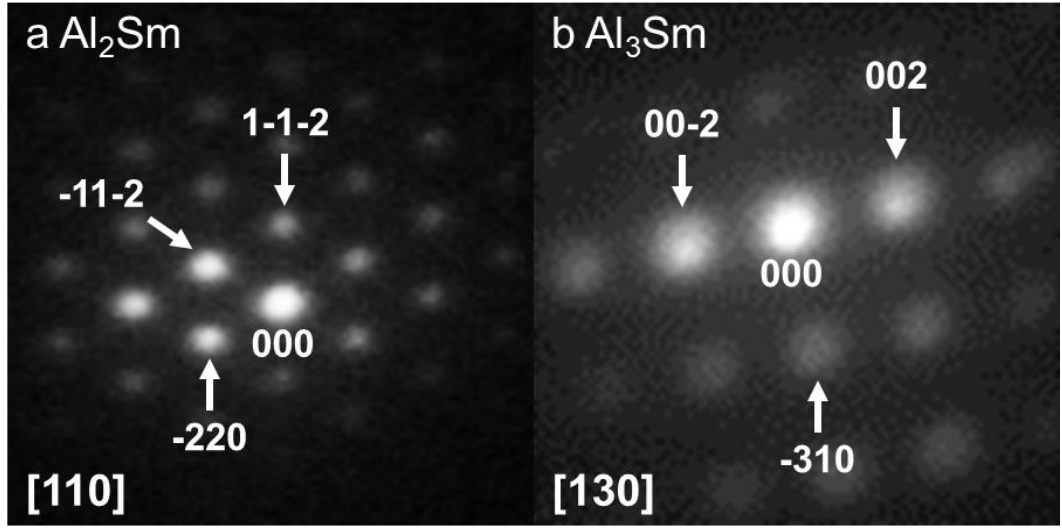

**Supplementary Figure 3.** Index of the nano-beam diffraction patterns of (a) Al$_2$Sm and (b) Al$_3$Sm.

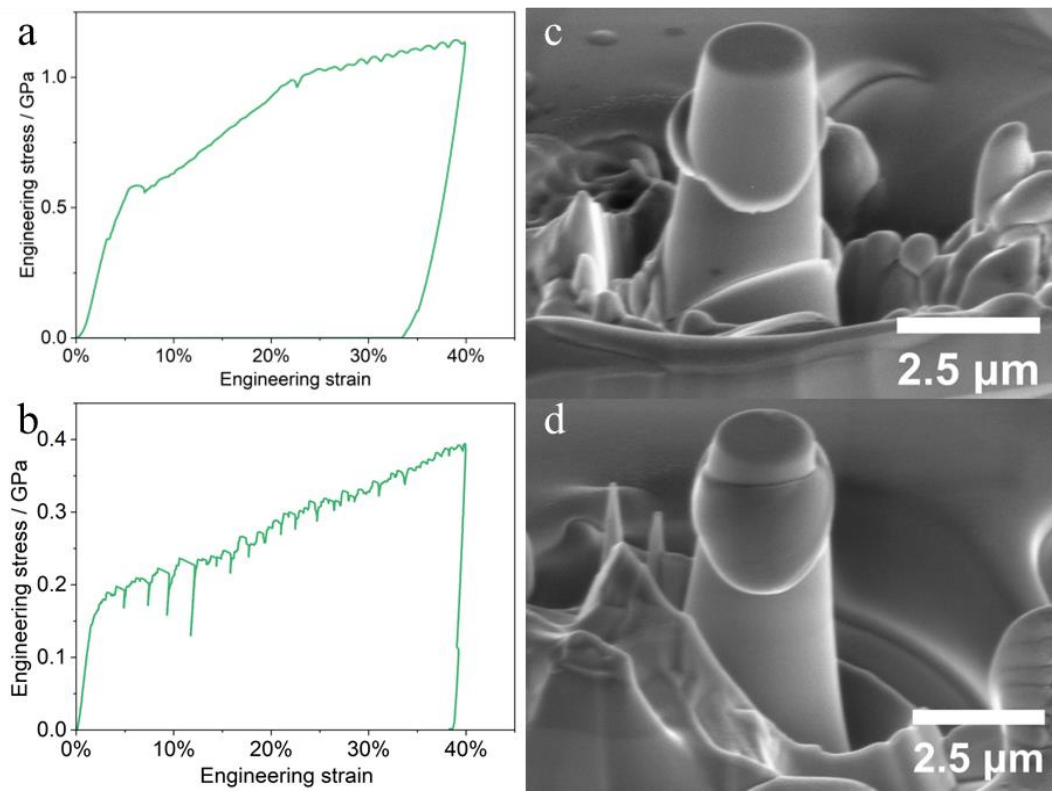

**Supplementary Figure 4.** (a) (b) Strain-stress curves for Al$_3$Sm micro-pillars experiencing 40 % engineering strain and (c) (d) related SEM images after compression.

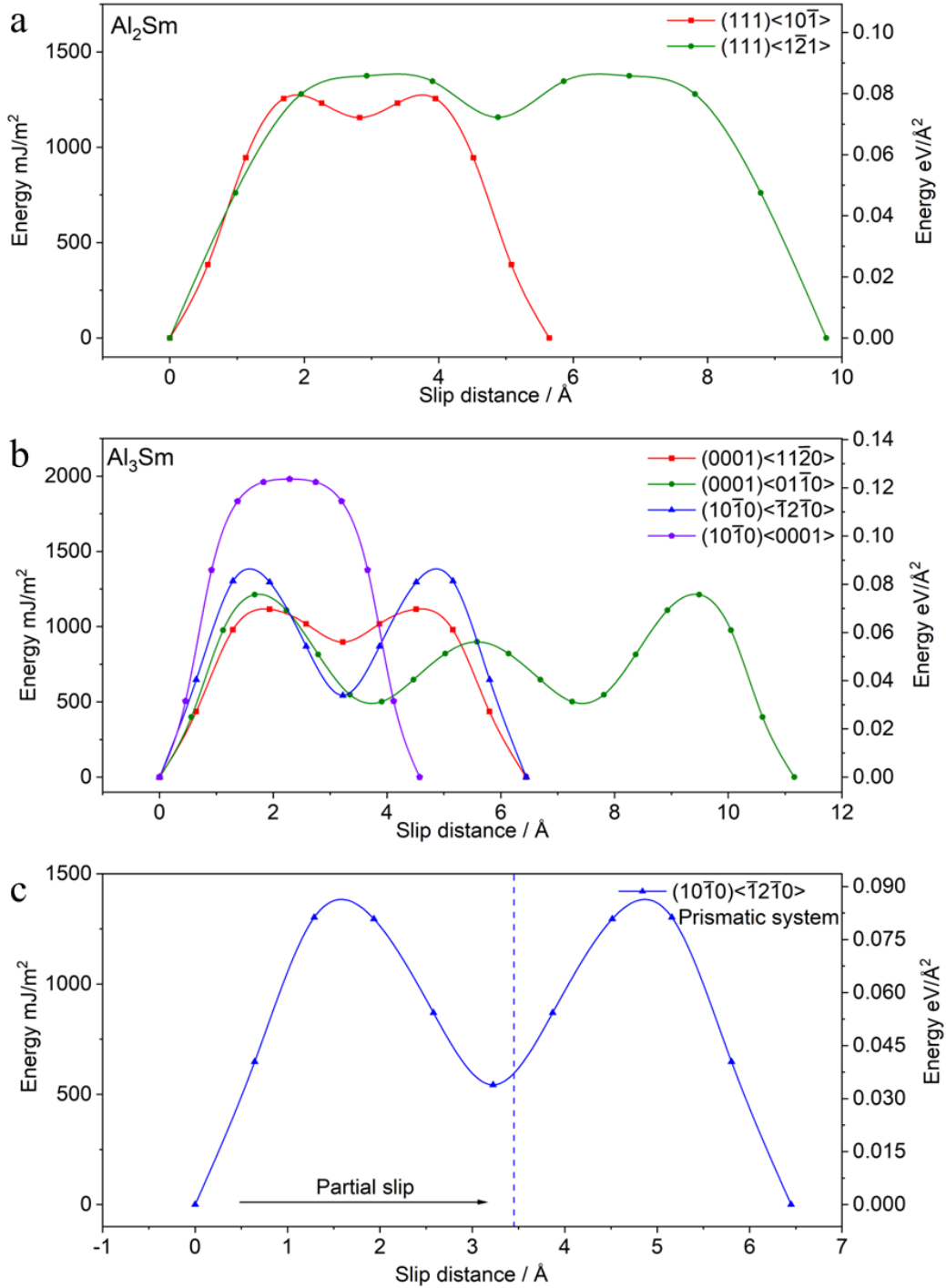

**Supplementary Figure 5.** Generalized Stacking Fault Energies (GSFE) on different potential energy surfaces of a rigid slip for (a) $Al_2Sm$ and (b) $Al_3Sm$; (c) partial slip on the prismatic slip system in $Al_3Sm$.

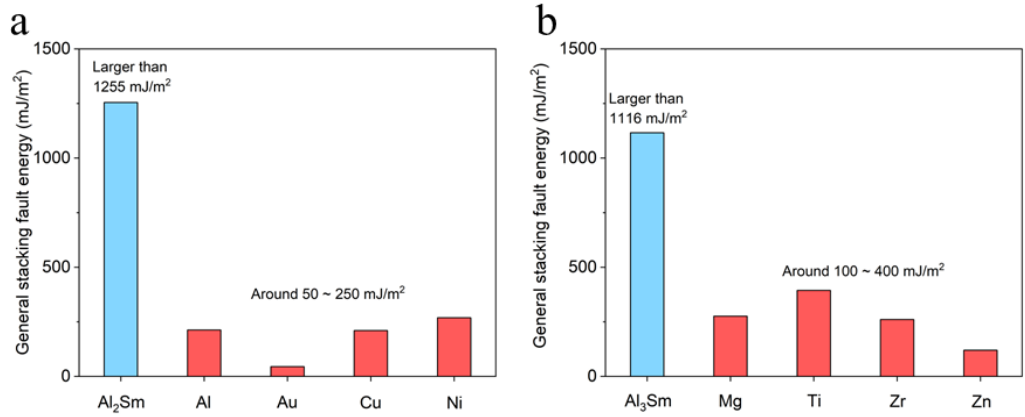

**Supplementary Figure 6.** Comparison of Generalized Stacking Fault Energy (GSFE) maxima between $Al_2Sm$, $Al_3Sm$, and materials with similar crystal structures (FCC and HCP when comparing to $Al_2Sm$ and $Al_3Sm$, respectively).

**Supplementary Table 1.** The energy maxima (in the unit of $mJ/m^2$) on different slip systems of $Al_2Sm$ and $Al_3Sm$ as well as of a number of other selected FCC and HCP materials.

| Slip system | $Al_2Sm$ | Al | Au | Cu | Ni |
|---|---|---|---|---|---|
| $(111)<10\bar{1}>$ | 1255 | 213[1,2] | 50[3] 45[4] | 210[5] | 269[5] |
| $(111)<1\bar{2}1>$ | 1374 |  | 83.6[6] |  |  |
|  |  |  |  |  |  |
| **Slip system** | **$Al_3Sm$** | **Mg** | **Ti** | **Zr** | **Zn** |
| $(0001)<11\bar{2}0>$ | 1116 | 276[7] | 394[8] | 261[8] | 120[8] |
| $(0001)<01\bar{1}0>$ | 1213 | 338[8] | 573[8] | 507[8] | 311[8] |
| $(10\bar{1}0)<\bar{1}2\bar{1}0>$ | 1303 |  |  |  |  |
| $(10\bar{1}0)<0001>$ | 1981 |  |  |  |  |

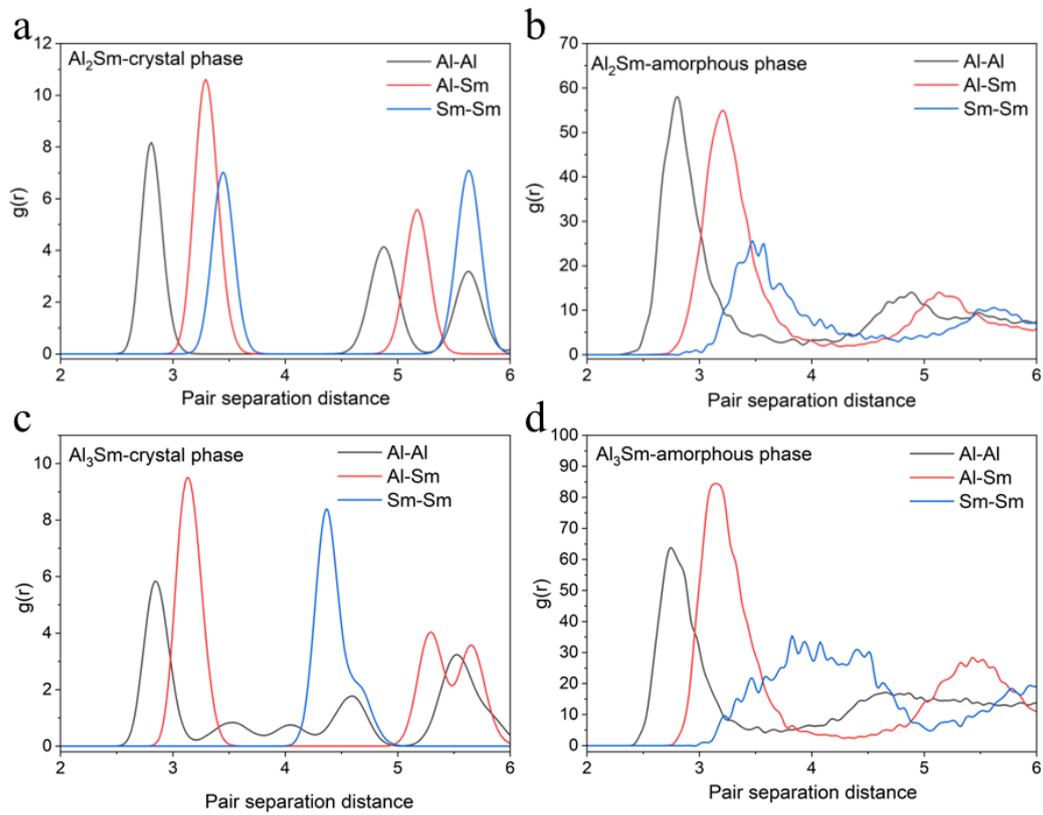

**Supplementary Fig. 7.** Radial Distribution Function of (a) crystal phase of Al$_2$Sm, (b) amorphous phase of Al$_2$Sm, (c) crystal phase of Al$_3$Sm and (d) amorphous phase of Al$_3$Sm.

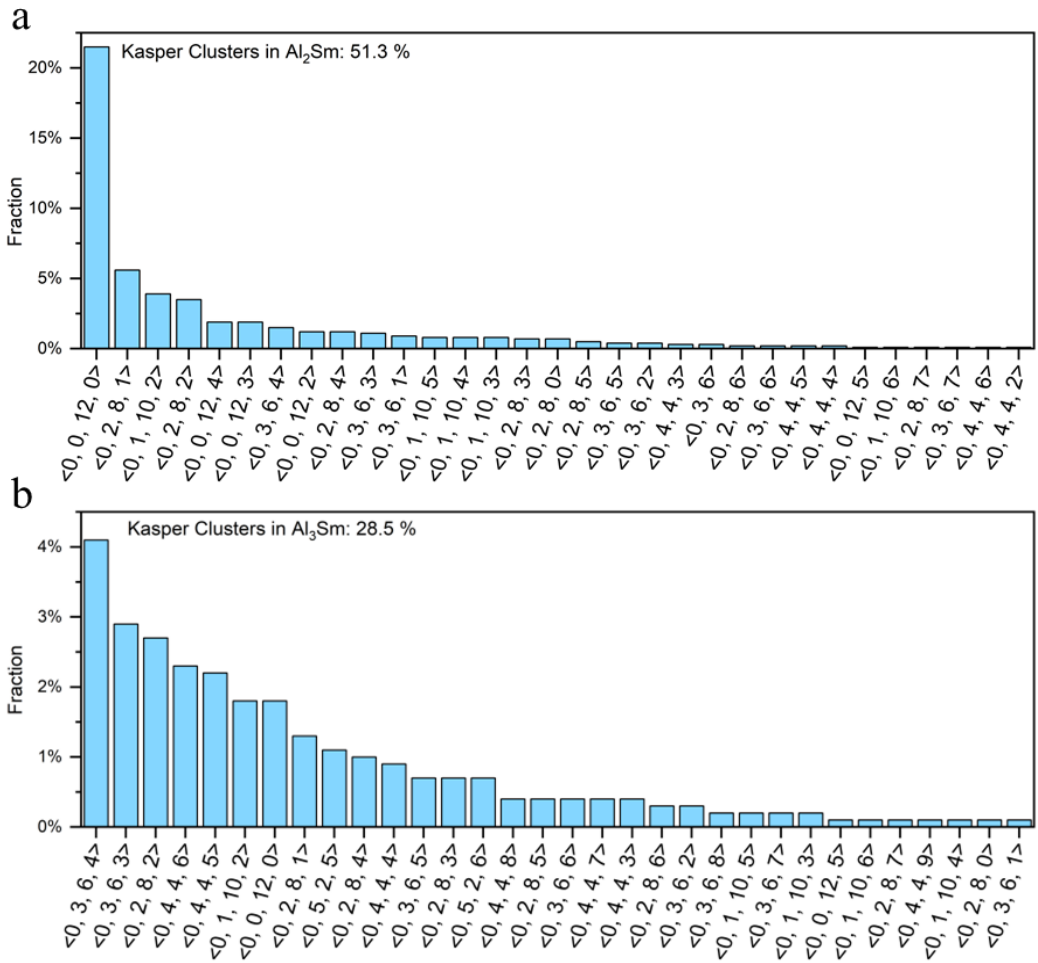

**Supplementary Fig. 8.** Voronoi indices of the Kasper clusters in amorphous phase in (a) Al$_2$Sm and (b) Al$_3$Sm.

**Supplementary Table 2.** Detailed error bar values (s.e.m) for Young's modulus and hardness of Al$_2$Sm and Al$_3$Sm in Fig. 5.

| Indent Displacement / nm | Young's Modulus / GPa | | Hardness / GPa | |
|---|---|---|---|---|
| | Al$_2$Sm | Al$_3$Sm | Al$_2$Sm | Al$_3$Sm |
| 106.60 | 162.34 ± 1.29 | | 10.54 ± 0.12 | |
| 108.68 | | 115.72 ± 2.63 | | 8.75 ± 0.05 |
| 125 | 162.39 ± 0.98 | 116.76 ± 2.35 | 10.39 ± 0.09 | 8.56 ± 0.05 |
| 150 | 162.07 ± 1.63 | 116.27 ± 2.35 | 10.05 ± 0.11 | 8.53 ± 0.08 |
| 175 | 161.36 ± 1.42 | 117.37 ± 2.25 | 9.91 ± 0.14 | 8.41 ± 0.06 |

| 200 | 161.65 ± 0.99 | 117.08 ± 2.49 | 9.81 ± 0.16 | 8.37 ± 0.06 |
| --- | --- | --- | --- | --- |
| 225 | 159.66 ± 1.32 | 118.04 ± 2.30 | 9.68 ± 0.15 | 8.22 ± 0.06 |
| 250 | 158.92 ± 1.02 | 117.07 ± 2.20 | 9.55 ± 0.18 | 8.19 ± 0.05 |
| 275 | 157.69 ± 0.80 | 116.70 ± 1.83 | 9.44 ± 0.19 | 8.11 ± 0.04 |
| 300 | 156.64 ± 0.77 | 117.02 ± 2.06 | 9.33 ± 0.19 | 7.96 ± 0.06 |
| 325 | 157.08 ± 0.57 | 116.29 ± 1.93 | 9.22 ± 0.16 | 7.93 ± 0.06 |
| 350 | | 116.20 ± 1.70 | | 7.88 ± 0.05 |
| 375 | | 115.72 ± 1.90 | | 7.87 ± 0.05 |

a Al$_2$Sm

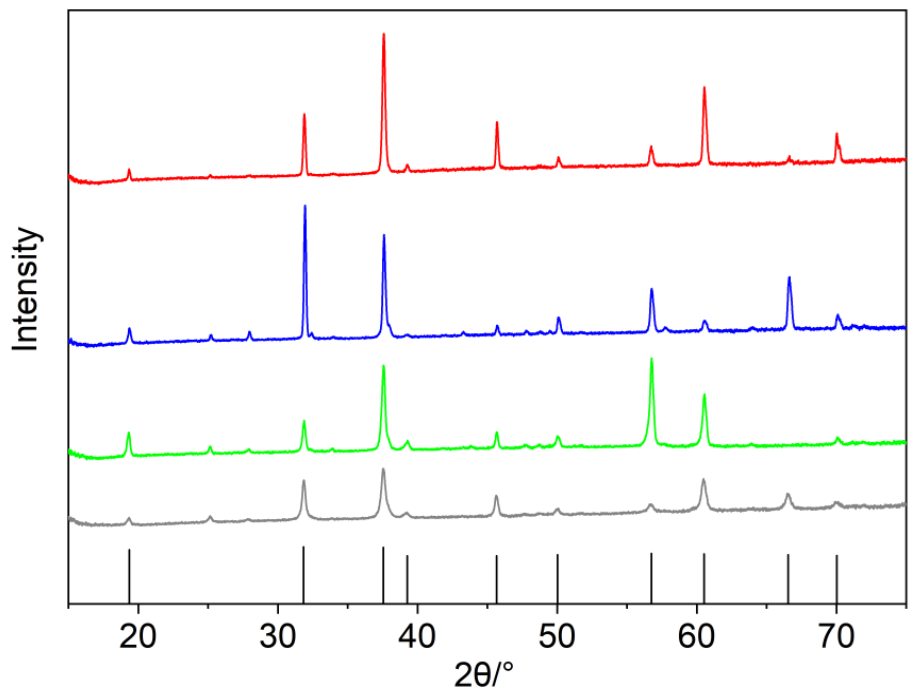

b Al$_3$Sm

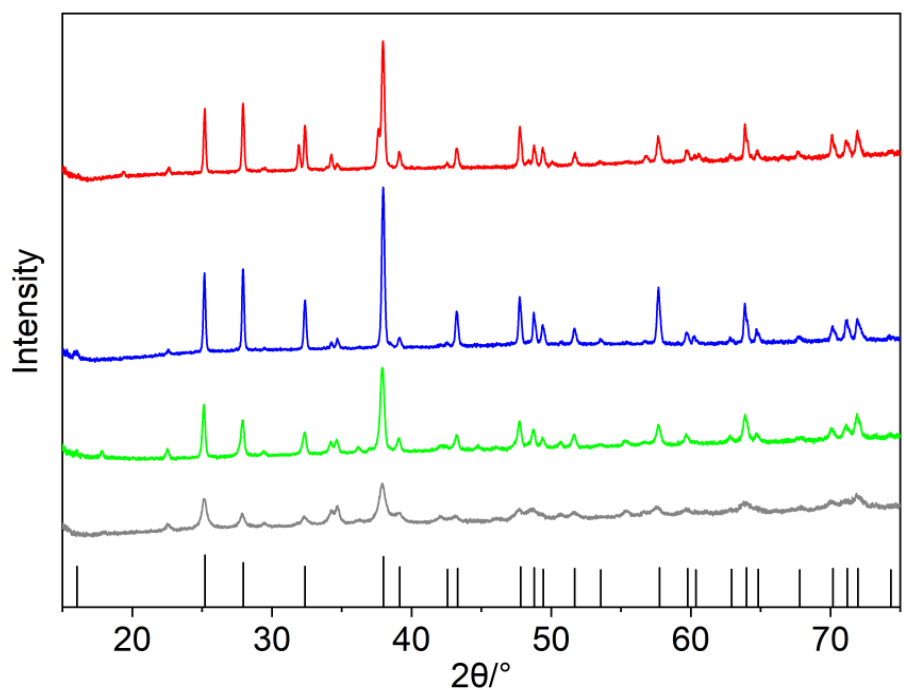

**Supplementary Figure 9.** XRD of the synthesized (a) Al$_2$Sm and (b) Al$_3$Sm crystalline samples. The different colors correspond to samples after arc-melting without annealing (black), and after annealing at 600℃ for 6 h (green), 12 h (blue),

and 50 h (red).

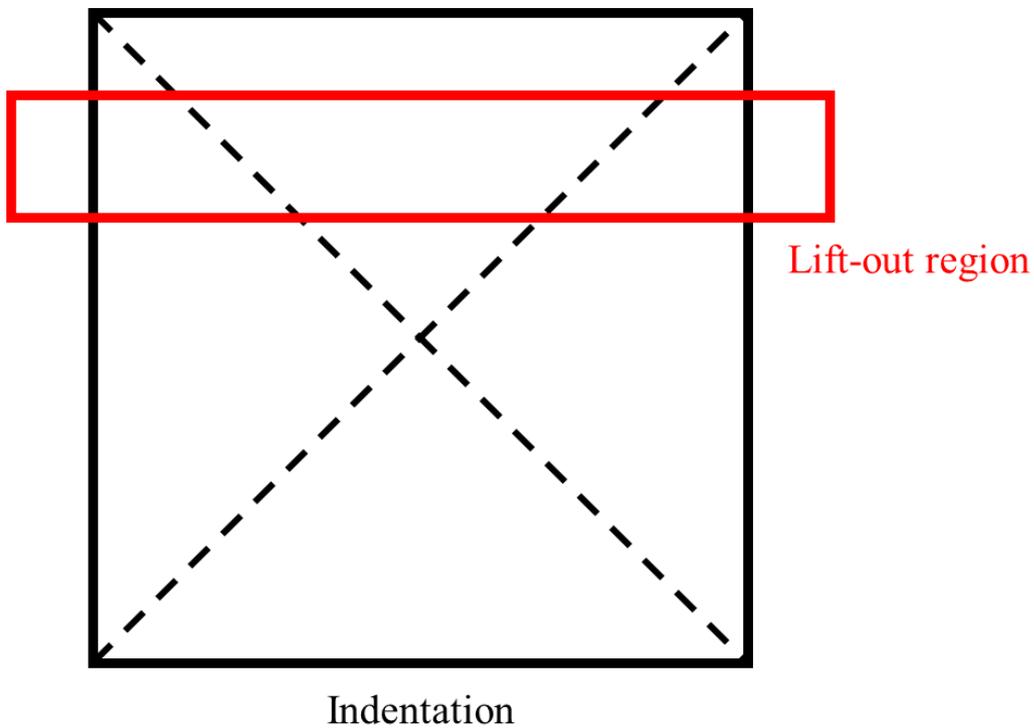

**Supplementary Figure 10.** Sketch of the lift-out region after Vicker's indentations.

### 3. References in Supplementary Information